\documentclass[11pt]{article}
\usepackage[margin=0.25in]{geometry}
\usepackage{authblk}

\usepackage{amsmath, amsfonts, amsthm}
\usepackage{threeparttable}
\usepackage{multirow}
\usepackage{hyperref}
\usepackage[table]{xcolor}
\usepackage{booktabs,tabulary}
\usepackage{siunitx}

\newcommand{\CP}{P }
\newcommand{\CPn}{\mathbb{P}_{n}}

\usepackage[english]{babel}
\newtheorem{theorem}{Theorem}
\newtheorem{lemma}{Lemma}
\begin{document}

\title{Weighted Mean Difference Statistics for Paired Data in Presence of Missing Values} 
\date{}
\author[1]{Yuntong Li}

\author[2,3]{Brent J. Shelton}
\author[3]{William St Clair} 

\author[2,3]{Heidi L. Weiss}

\author[3]{John L. Villano}

\author[4]{Arnold J. Stromberg}

\author[2,3,4]{Chi Wang}

\author[2,3]{Li Chen \thanks{Corresponding Author. Email: lichenuky@uky.edu}}

\affil[1]{AbbVie Inc., Illinois, Unites States}

\affil[2]{Biostatistics and Bioinformatics Shared Resource Facility, Markey Cancer Center, University of Kentucky, Kentucky, Unites States}
\affil[3]{Department of Internal Medicine, University of Kentucky, Kentucky, Unites States}
\affil[4]{Department of Statistics, University of Kentucky, Kentucky, Unites States}

\begin{titlepage}
\maketitle
\end{titlepage}


\abstract{Missing data is a common issue in many biomedical studies. Under a paired design, some subjects may have missing values in either one or both of the conditions due to loss of follow-up, insufficient biological samples, etc. Such partially paired data complicate statistical comparison of the distribution of the variable of interest between the two conditions. In this paper, we propose a general class of test statistics based on the difference in weighted sample means without imposing any distributional or model assumption. An optimal weight is derived for this class of tests. Simulation studies show that  our proposed test with the optimal weight performs well and outperforms existing methods in practical situations. Two cancer biomarker studies are provided for illustration.

}


\maketitle


\section{Introduction}\label{intro}
Comparing the distribution of a variable of interest between two experimental conditions/groups is a common problem in many biomedical studies. A paired design is frequently used because it can reduce the variability in the observations due to differences between individuals. However, in practice, the data are often partially paired, where only a proportion of individuals have measurements under both conditions while other individuals have missing measurements in either one or both of the conditions. For example, a study enrolled a total of 18 prostate cancer patients who underwent surgery. The study longitudinally measured three biomarkers, including PSA, CRP and IL8, at baseline (prior to surgery), 3-month and 6-month follow-ups with a goal of comparing those biomarker levels before and after  surgery.  Due to missed visits or loss of follow-up, 61\% of patients had missing values at 3-month and 50\% had missing values at 6-month follow-up. As another example, a breast cancer biomarker study was conducted to compare expression of five proteins of interest between primary and recurrent tumors \cite{he2019potential}. Tissue blocks of paired primary and recurrent tumors from 22 breast cancer patients were selected as part of a tissue microarray construction. Immunohistochemistry (IHC) was performed to measure the protein expressions. A 0 to 3 intensity score was assigned to each tissue core by a pathologist. Missing values were present in primary and/or recurrent tumor tissue samples due to the lack of remaining tissue, lack of invasive carcinoma cells or difficulty in interpreting tissues. The missing proportion in primary tumor tissue samples ranged from 5\% to 14\%, and that in recurrent tumor tissue samples ranged from 0\% to 19\%.

Let $(Y_1, Y_2)$ denote the paired variables from groups 1 and 2, and $(Y_{1,i}, Y_{2,i})$, $i = 1, \ldots, n,$ be a simple random sample of $(Y_1, Y_2)$. For partially paired data, suppose $(Y_1, Y_2)$ are both observed in individuals $1, \ldots, n_0$, $Y_1$  alone is  observed in individuals $n_0+1, \ldots, n_0+n_1$, and $Y_2$ alone is observed in individuals $n_0+n_1+1, \ldots, n_0+n_1+n_2$. In this paper, we assume missingness of $Y_g$ is independent of $Y_g$ for $g=1,2$, i.e. missing completely at random (MCAR) \cite{little2002statistical}. The observed data can be divided into the following four blocks: group 1 data from complete pairs $Y_1^C = (Y_{1,1}, \ldots, Y_{1,n_0})$,   group 1 data from incomplete pairs $Y_1^I = (Y_{1,n_0+1}, \ldots, Y_{1,n_0+n_1})$, group 2 data from complete pairs $Y_2^C = (Y_{2,1}, \ldots, Y_{2,n_0})$, and group 2 data from incomplete pairs $Y_2^I = (Y_{2,n_0+n_1+1}, \ldots, Y_{2,n_0+n_1+n_2})$. The data structure is illustrated below.

\vspace{-0.5cm}
\begin{center}
	\begin{table*}[h!]%
		\centering
		\caption{Data Structure}%
		\begin{tabular}{cccc}
		Both $Y_1$ and $Y_2$ are observed & $Y_1$ alone is observed & $Y_2$ alone is observed  & Both $Y_1$ and $Y_2$ are missing\\
		$Y_{1,1}, \ldots, Y_{1,n_0}$ & $Y_{1,n_0+1}, \ldots, Y_{1,n_0+n_1}$ & &\\
		$Y_{2,1}, \ldots, Y_{2n_0}$  & & $Y_{2,n_0+n_1+1}, \ldots, Y_{2,n_0+n_1+n_2}$ &
		\end{tabular}
		\end{table*}
		\end{center}
\vspace{-0.5cm}

Naive methods to handle partially paired data are either only using data of complete pairs  or imputing the missing values by sample mean or median and then applying standard statistical tests. However, only using data of complete pairs is inefficient because it discards part of the data. Imputing the missing values by sample mean will underestimate the variances of estimates, and lead to erroneous results\cite{little2002statistical}. To fully utilize all available data, the mixed models, including the linear mixed model and cumulative link mixed model, can be applied. However, those models require data normality assumption and distributional assumption on random effects. In addition to the naive methods and mixed models, statistical tests specifically designed for partially paired data, including rank-based and mean difference-based tests, have been proposed. The rank-based tests considered weighted linear combinations of Wilcoxon signed rank and Wilcoxon rank-sum tests. Dubnicka {\it et al.} (2002) \cite{dubnicka2002rank} combined a Wilcoxon signed rank test statistic on complete pairs, i.e. comparing $Y_1^C$ and $Y_2^C$, and a Wilcoxon rank-sum test statistic on incomplete pairs, i.e. comparing $Y_1^I$ and $Y_2^I$. Brunner and Puri (1996) \cite{brunner2002nonparametric}, Konietschke {\it et al.} (2012) \cite{konietschke2012ranking} and Fong {\it et al.} (2018)\cite{fong2017rank} considered all possible comparisons of blocks between groups, including $Y_1^C$ and $Y_2^C$, $Y_1^C$ and $Y_2^I$, $Y_1^I$ and $Y_2^C$, and $Y_1^I$ and $Y_2^I$. One shortcoming of rank-based tests is that they might not be sensitive to the magnitude of the difference between groups. In contrast, mean difference-based tests took into account the absolute values of observations and could be more powerful. Bhoj (1978) \cite{bhoj1978testing}  considered a weighted linear combination of the paired and unpaired t-test statistics under the assumption of bivariate normal distribution.  Mart{\'\i}nez-Camblor {\it et al.} (2013) \cite{martinez2013hypothesis} proposed a test based on the difference in sample means between groups and obtained asymptotic normal distribution of the test without any distribution assumption.

In this paper, we propose a general class of test statistics based on the difference in weighted sample means of the two groups for partially paired data. This general class of test statistics include the test statistics proposed in Bhoj (1978) \cite{bhoj1978testing}  and Mart{\'\i}nez-Camblor {\it et al.} (2013) \cite{martinez2013hypothesis} as special cases. We further derive an optimal  weight for this class of tests. Our tests are constructed without imposing any distributional or model assumption. Therefore, they are applicable to continuous data with an arbitrary distribution as well as discrete interval data.  We evaluate the performance of the proposed test with optimal weight and compare it with several existing  tests by using simulation studies and real data analyses.

\section{Methods}\label{sec2}
\subsection{Weighted mean difference test statistics}
 Let $\mu_1$= E$Y_{1}$ and $\mu_2$= E$Y_{2}$. To test whether there is a difference in distribution between paired outcomes $Y_{1}$ and  $Y_{2}$, we focus on examining whether there is a mean difference, i.e. $H_0$: $D=\mu_1-\mu_2=0$  vs.  $H_1$: $D\ne0$. We are going to first propose a general class of estimators for $\mu_1-\mu_2$, and then construct Wald tests for  $H_0$ based on those estimators. Specifically, we consider the following class of estimators 
\begin{equation}\label{eq:D}
\hat{D}(\hat{w}_1, \hat{w}_2) = \left\{\hat{w}_1 \bar{Y}_{1}^C +(1-\hat{w}_1)\bar{Y}_{1}^I\right \}- \left \{\hat{w}_2 \bar{Y}_{2}^C +(1-\hat{w}_2)\bar{Y}_{2}^I\right \},
\end{equation}
where $\bar{Y}_{g}^C$, $\bar{Y}_{g}^I$ ($g=1, 2$) are the means for complete pair data $Y_{g}^C$ and incomplete pair data $Y_{g}^I$, respectively, and $\hat{w}_1 $ and $\hat{w}_2$ are random weights in [0, 1].  The first item on the right hand side of equation (\ref{eq:D}) is a weighted sample mean of group 1, and the second item is a weighted sample mean of group 2. Therefore, $\hat{D}(\hat{w}_1, \hat{w}_2)$ quantifies the difference in weighted sample means  between  two groups. As a special case, when 
$\hat{w}_1 = n_0 / (n_0+n_1)$ and $\hat{w}_2 = n_0 / (n_0+n_2)$,
the $\hat{D}(\hat{w}_1, \hat{w}_2)$ reduces to the difference in sample mean. As another special case, when $\hat{w}_1 = \hat{w}_2 =1$, the $\hat{D}(\hat{w}_1, \hat{w}_2)$ reduces to the difference in sample mean only using complete pairs.

Our proposed class of estimators provide a general weighting mechanism to take into account two specific characteristics of the data. First, for observations from the same group, those from complete pairs and from incomplete pairs are given different weights because they contain different amount of information. Second, observations from different groups are given different weights because different groups may have different missing proportions and different variances. In the following Theorem \ref{theorem2}, we establish the asymptotic normality of $\hat{D}(\hat{w}_1, \hat{w}_2)$. The proof of this theorem is provided in Appendix A.

\begin{theorem}[]\label{theorem2}
	Suppose there exist deterministic weights $0 \leq w_1 \leq 1$ and $0 \leq w_2 \leq 1$ such that $\sqrt{n}(\hat{w}_1-w_1)$ and $\sqrt{n}(\hat{w}_2-w_2)$ are $O_p(1)$. Under the MCAR assumption, 
	\begin{align*}
		\sqrt{n}\left\{\hat{D}(\hat{w}_1, \hat{w}_2) - (\mu_1 - \mu_2)\right\} \rightarrow N\left(0, \sigma_D^2(w_1, w_2)\right). 
	\end{align*}
\noindent When there are missing values in both groups,
\begin{eqnarray}\label{eq:sigmaD}
	\sigma_D^2(w_1, w_2) &=&
	\left\{\frac{w_1^2}{p_{12}} + \frac{(1-w_1)^2}{p_1 - p_{12}}\right\}\sigma_1^2 +  \left\{\frac{w_2^2}{p_{12}} + \frac{(1-w_2)^2}{p_2 - p_{12}}\right\}\sigma_2^2 - \frac{2w_1w_2\rho \sigma_1 \sigma_2}{p_{12}},
\end{eqnarray}
where $p_g$ is the probability that $Y_g$ is observed, $p_{12}$ is the probability that both $Y_1$ and $Y_2$ are observed, $\sigma_g^2$ is the variance of $Y_{g}$, and $\rho$ is the correlation coefficient of $Y_{1}$ and $Y_{2}$. When missing values only present in group 1,  the $\sigma_D^2(w_1, w_2)$ is given by equation (\ref{eq:sigmaD}) with $w_1=1$, $p_2=1$, $p_{12} = p_1$ and $0/0=0$. When missing values only present in group 2,  the $\sigma_D^2(w_1, w_2)$ is given by equation (\ref{eq:sigmaD}) with $w_2=1$, $p_1=1$, $p_{12} = p_2$ and $0/0=0$. 
\end{theorem}

Based on $\hat{D}(\hat{w}_1, \hat{w}_2)$, we construct the following class of weighted mean difference (WMD) test statistics to assess the  hypothesis $H_0$ versus $H_1$:
\[T(\hat{w}_1, \hat{w}_2) = \frac{\sqrt{n}\hat{D}(\hat{w}_1, \hat{w}_2)}{\hat{\sigma}_D (\hat{w}_1, \hat{w}_2)}, \]
where $\hat{\sigma}_D(\hat{w}_1, \hat{w}_2)$ is an estimator of $\sigma_D(w_1, w_2)$ by plugging in $\hat{w}_1$, $\hat{w}_2$ as well as the following estimators into equation (\ref{eq:sigmaD}): $\hat{p}_g = (n_0+n_g)/n$ is the non-missing proportion and $\hat{\sigma}_g$ is the sample standard deviation for group $g$, $g = 1, 2$, $\hat{p}_{12} = n_0/n$ is the proportin of complete pairs, and $\hat{\rho}$ is the sample correlation coefficient using complete pairs. Based on Theorem \ref{theorem2}, $T(\hat{w}_1, \hat{w}_2)$ converges to a standard normal distribution, $N(0,1)$, under $H_0$. Therefore, the critical region of WMD test statistics is $\{|T(\hat{w}_1, \hat{w}_2)| > z_{1-\alpha/2}\}$, where $\alpha$ is the significance level. When the sample size is small, $\hat{\sigma}_D (\hat{w}_1, \hat{w}_2)$ can underestimate the standard error of $\hat{D}(\hat{w}_1, \hat{w}_2)$, which leads to an inflated type I error rate of WMD tests. Therefore, we recommend using a bootstrap-based estimator for the standard error of $\hat{D}(\hat{w}_1, \hat{w}_2)$ in the calculation of WMD tests.

\subsection{Power and optimal weight of WMD tests}
According to Theorem 1, under $H_1$, $T(\hat{w}_1, \hat{w}_2)$ converges to $N((\mu_1 - \mu_2)/\sigma_D(w_1, w_2), 1)$. Therefore, the power of WMD test statistics is $1- \Phi[z_{1-\alpha/2} - (\mu_1 - \mu_2)/\sigma_D(w_1, w_2)] + \Phi[-z_{1-\alpha/2} - (\mu_1 - \mu_2)/\sigma_D(w_1, w_2)]$. Because the power is a decreasing function of $\sigma_D(w_1, w_2)$, it is maximized when $\sigma_D(w_1, w_2)$ is minimized. We present here the optimal weight that minimizes  $\sigma_D(w_1, w_2)$ with a detailed derivation provided in Appendix B. 

When there are missing values in both groups, the optimal weight is
\begin{eqnarray}\label{eq:weight}
(w_1^{\textrm{opt}}, w_2^{\textrm{opt}}) &=& \left\{
\begin{split}
(w_1^*, w_2^*) \hspace{1cm} & \hspace{0.3cm}  \textrm{if } 0 \leq w_1^* \leq 1, 0 \leq w_2^* \leq 1,\\
\underset{(w_1, w_2) \in \mathcal{A}}{\textrm{argmin}} \sigma_D^2(w_1, w_2) & \hspace{0.3cm}  \textrm{otherwise}, \\
\end{split}\right.
\end{eqnarray}
with 
\begin{eqnarray*}
w_{1}^{*} =  \frac{p_2p_{12}\sigma_1^2 + \rho\sigma_1\sigma_2p_{12}(p_1-p_{12})}{p_1p_2\sigma_1^2 - \rho^2\sigma_1^2(p_1-p_{12})(p_2-p_{12})}, \hspace{0.4cm} w_{2}^{*} =  \frac{p_1p_{12}\sigma_2^2 + \rho\sigma_1\sigma_2p_{12}(p_2-p_{12})}{p_1p_2\sigma_2^2 - \rho^2\sigma_2^2(p_1-p_{12})(p_2-p_{12})},\\
\end{eqnarray*}
and $\mathcal{A}$ is a collection of  four points
$\mathcal{A}  =   \{ (0, h_2(0)), (1, g(h_2(1))), 
(h_1(0), 0), (g(h_1(1),1)\},$ $h_1(x) =  (p_1\sigma_1)^{-1} (p_{12}\sigma_1+x\rho\sigma_2(p_1-p_{12}))$, $h_2(x) =  (p_2\sigma_2)^{-1} (p_{12}\sigma_2+x\rho\sigma_2(p_2-p_{12}))$, and $g(z) \textrm{ equals }0 \textrm{ if } z<0$; $\textrm{ equals }z \textrm{ if } 0\leq z \leq 1$; and $g(z) =1 \textrm{ if } z>1$.

When missing values only present in group 1, we have $p_2=1$, $p_{12}=p_1$ and $w_1=1$.  The optimal weight for $w_2$ is $w_2^{\textrm{opt}} = g(p_{1} +\rho\sigma_1\sigma_2^{-1} (1-p_{1}))$. When missing values only present in group 2, we have $p_1=1$, $p_{12}=p_2$ and $w_2=1$. The optimal weight for $w_1$ is $w_1^{\textrm{opt}} = g(p_{2} +\rho\sigma_1^{-1}\sigma_2 (1-p_{2}))$.

The optimal weight is estimated by replacing $p_1, p_2, p_{12}, \sigma_1, \sigma_2$ and  $\rho$ with their estimators as described in Section 2.1. The WMD test with this estimated optimal weight is referred to as $\textrm{WMD}_{\textrm{o}}$.

\subsection{Relationships to existing methods}
Our proposed test is constructed by calculating the difference in weighted sample means between the two groups. When we set
$$\hat{w}_1 = \frac{n_0}{n_0+n_1}, \hspace{0.3cm} \hat{w}_2 = \frac{n_0}{n_0+n_2},$$
the weight for each observation in the first group becomes $1/(n_0 + n_1)$, and that for each observation in the second group becomes $1/(n_0 + n_2)$.
The $\hat{D}$ reduces to 
$$\hat{D} = \bar{Y}_1 - \bar{Y}_2,$$
where $\bar{Y}_g$ is the sample mean of the $g$th group. The $\hat{D}$ is the simple mean difference between groups, which is referred to as $\rm{WMD}_{\rm {s}}$. The corresponding asymptotic variance formula reduces to 
$$\frac{\sigma_1^2}{p_1} + \frac{\sigma_2^2}{p_2} - \frac{2\rho \sigma_1\sigma_2p_{12}}{p_1p_2}.$$
The $\rm{WMD}_{\rm {s}}$ is asymptotically equivalent to the $\mathcal{M}_n$ test \cite{martinez2013hypothesis}.
Note that the $\mathcal{M}_n$ test assigns equal weight to each observation from the same group. However, the complete pair and incomplete pair may contain different amount of information. By introducing a weighted sample mean, out test allows different weights to complete and incomplete data, and thus could improve the power of the test. In addition, by introducing the weight, our test can also be considered as a weighted combination of paired and unpaired $t$ statistics. Let 
$$\hat{w}_1 = \hat{w}_2 = \frac{\frac{\lambda}{s\left/\sqrt{n_0}\right.}}{\frac{\lambda}{s\left/\sqrt{n_0}\right.} + \frac{1-\lambda}{s_1\left/\sqrt{1/n_1 + 1/n_2}\right.}}.$$
Our test statistic reduces to the test proposed by Bhoj (1978) \cite{bhoj1978testing}. Bhoj (1978) \cite{bhoj1978testing} assumed that data are normally distributed and calculated p-values based on a $t$-distribution. As we do not require data to be normally distributed,  our p-value is obtained using either normal approximation or bootstrap. 


\section{Simulation Studies}\label{sec3}
\subsection{Continuous data}
We first evaluated the performance of the proposed WMD test with the optimal weight ($\rm{WMD}_{\rm {o}}$) for analyzing continuous data, and compared to the paired $t$-test on complete pairs ($\rm{t}_{\rm {cp}}$), paired $t$-test with imputation ($\rm{t}_{\rm {im}}$), linear mixed model (LMM), Wilcoxon signed-rank test on complete pairs ($\rm{W}_{\rm {cp}}$), Wilcoxon signed-rank test with imputation ($\rm{W}_{\rm{im}}$), and $\rm{WMD}_{\rm {s}}$ which is a test based on the difference in sample means between groups and is asymptotically equivalent to the $\mathcal{M}_n$ test \cite{martinez2013hypothesis}. For $\rm{t}_{\rm {im}}$ and $\rm{W}_{\rm{im}}$, missing values were imputed by median of each group. For $\rm {LMM}$, we used the \textit{nlme} R-package\cite{nlme2019} with the default restricted maximum likelihood setting. To mimic real data situation, we simulated data based on complete pairs of PSA data from the prostate cancer study described in section 4.1. We randomly sampled $n$ subjects from the PSA data with replacement. For each subject, both the baseline and 6-month follow-up measurements were selected, which maintained the within-subject correlation in the real data. Because the baseline and 6-month measurements had large differences, all methods would yield powers close to one for detecting such difference. To better investigate the difference among methods,  we reduced the difference between those two measurements by subtracting a value of 6 from the baseline measurement. We considered two scenarios of the correlation between baseline and 6-month measurements. The first scenario directly adopted the correlation from the real data, which had a Spearman correlation coefficient $R$=0.60. The second scenario was a weaker correlation with $R$=0.33, which was generated by adding a random normal error term to V00 measurement. We considered two missing scenarios:  missing only in group 2 with 50\% missing in group2 and missing in both groups with 50\% missing in both groups. We considered the sample size $n$ equal to 20, 30, 50 or 100. Under each sample size, we replicated the simulation 1000 times.

Our first set of simulations evaluated the type I error rate of our proposed method as well as other methods. To generate data under the null hypothesis, we randomly rearranged the labels of the paired baseline and 6-month measurements so that distributions of simulated data under the two time points were expected to be the same.   Table~ \ref{psa_type1_t06} shows the type I error rate for each method at a given significant level $\alpha=0.05$.
 $\rm{WMD_o}$, $\rm{WMD}_{\rm {s}}$ and $\rm{W}_{\rm {cp}}$ were able to control the type I error rate close to the nominal level of 0.05 under all sample sizes. $\rm{t}_{\rm {cp}}$ and LMM were too conservative when $R$=0.60 and the sample size was small. But their type I error rates became close to the nominal level as the sample size increased. $\rm{t}_{\rm {im}}$ had inflated type I error rate when $R$=0.33, and  $\rm{W_{im}}$ had substantially inflated type I error rate under most scenarios, which were likely due to the reduced data variation after imputation.

\begin{center}
	\begin{table}[h!]%
		\centering
		\caption{Simulation results for type I error rates based on continuous data\label{psa_type1_t06}}%
		\begin{tabular*}{500pt}{@{\extracolsep\fill}lccccccccc@{\extracolsep\fill}}
			\hline
		\textbf{Corr} &\textbf{Missing} &	\textbf{n} & $\mathbf{t_{cp}}$  & $\mathbf{t_{im}}$  & $\mathbf{LMM}$ & $\mathbf{W_{cp}}$  & $\mathbf{W_{im}}$  & $\mathbf{WMD_{s}}$  & $\mathbf{WMD_o}$ \\
			\hline
			0.60 & (0\%, 50\%) & 20 & 0.01 & 0.04 & 0.02 & 0.03 & 0.08 & 0.07 & 0.07 \\ 
			& & 30 & 0.02 & 0.04 & 0.04 & 0.05 & 0.08 & 0.06 & 0.07 \\ 
			& & 50 & 0.05 & 0.05 & 0.05 & 0.05 & 0.07 & 0.07 & 0.07 \\ 
			& & 100 & 0.05 & 0.06 & 0.05 & 0.05 & 0.06 & 0.06 & 0.06 \\ 
			 & (50\%, 50\%) & 20 & 0.02 & 0.03 & 0.02 & 0.02 & 0.18 & 0.04 & 0.07 \\ 
			 & & 30 & 0.01 & 0.03 & 0.01 & 0.03 & 0.15 & 0.04 & 0.07 \\ 
			 & & 50 & 0.02 & 0.04 & 0.03 & 0.04 & 0.09 & 0.04 & 0.05 \\ 
			 & & 100 & 0.04 & 0.05 & 0.04 & 0.04 & 0.04 & 0.05 & 0.05 \\ 
			0.32 & (0\%, 50\%) & 20 & 0.03 & 0.07 & 0.04 & 0.04 & 0.12 & 0.06 & 0.07 \\ 
			& & 30 & 0.04 & 0.06 & 0.04 & 0.04 & 0.12 & 0.06 & 0.06 \\ 
			& & 50 & 0.04 & 0.07 & 0.05 & 0.05 & 0.12 & 0.06 & 0.07 \\
			& & 100 & 0.07 & 0.08 & 0.07 & 0.07 & 0.15 & 0.07 & 0.06 \\ 
			& (50\%, 50\%) & 20 & 0.04 & 0.18 & 0.03 & 0.02 & 0.34 & 0.06 & 0.07 \\
			& & 30 & 0.03 & 0.19 & 0.04 & 0.03 & 0.39 & 0.06 & 0.06 \\
			& & 50 & 0.04 & 0.16 & 0.05 & 0.05 & 0.43 & 0.06 & 0.06 \\ 
			& & 100 & 0.05 & 0.15 & 0.06 & 0.05 & 0.48 & 0.06 & 0.05 \\ 
			\hline
		\end{tabular*}
		\begin{tablenotes}
			\item Each entry was based on 1000 simulation replicates. $\rm{WMD_{s}}$  and $\rm{WMD_o}$ were calculated based on 1500 bootstrap samples. Significant level $\alpha$ was set to 0.05. 
		\end{tablenotes}
	\end{table}
\end{center}

Our next set of simulations evaluated and compared the power of different methods. Because imputation methods had inflated type I error rates, we excluded them from the power comparison. We generated data following a similar procedure as the simulations for type I error rate evaluation, except that the labels of the paired baseline and 6-month measurements from the PSA data were maintained.  As shown in Table \ref{psa_power_t06}, $\rm {WMD_o}$ performed the best across all scenarios. $\rm {WMD_o}$ had much higher power than $\rm{t}_{\rm {cp}}$ and $\rm{W}_{\rm {cp}}$, since those two methods only used the portion of complete pairs data. $\rm {WMD_o}$ also had much higher power than LMM for small sample sizes, likely due to LMM's requirement of data normality that was not satisfied in the real data (P<0.001 based on Shapiro-Wilk test). In addition, the power of $\rm {WMD_o}$ was higher than that of $\rm {WMD_s}$, especially for $R=0.6$.

\begin{center}
	\begin{table}[h!]%
		\centering
		\caption{Simulation results for power comparison based on continuous data\label{psa_power_t06}}%
		\begin{tabular*}{500pt}{@{\extracolsep\fill}lccccccccc@{\extracolsep\fill}}
			\hline
			\textbf{Corr} &\textbf{Missing} &	\textbf{n} & $\mathbf{t_{cp}}$   & $\mathbf{LMM}$ & $\mathbf{W_{cp}}$    & $\mathbf{WMD_{s}}$  & $\mathbf{WMD_o}$ \\
			\hline
			0.60 & (0\%, 50\%) & 20 & 0.07 & 0.04 & 0.08 & 0.55 & 0.68 \\ 
			& & 30 & 0.21 & 0.14 & 0.16 & 0.78 & 0.85 \\
			& & 50 &  0.49 & 0.50 & 0.23 & 0.96 & 0.98 \\ 
			& & 100 & 0.94 & 0.99 & 0.45 & 1.00 & 1.00 \\ 
			& (50\%, 50\%) & 20 & 0.01 & 0.03 & 0.02 & 0.20 & 0.44 \\ 
			& & 30 & 0.05 & 0.16 & 0.07 & 0.35 & 0.56 \\ 
			& & 50 & 0.13 & 0.44 & 0.10 & 0.56 & 0.72 \\ 
			& & 100 & 0.51 & 0.93 & 0.22 & 0.92 & 0.96 \\
			0.32 & (0\%, 50\%) & 20 & 0.16 & 0.17 & 0.18 & 0.41 & 0.41 \\
			& & 30 & 0.30 & 0.32 & 0.29 & 0.56 & 0.56 \\ 
			& & 50 & 0.46 & 0.53 & 0.39 & 0.68 & 0.68 \\
			& & 100 & 0.71 & 0.75 & 0.53 & 0.84 & 0.85 \\ 
			& (50\%, 50\%) & 20 & 0.08 & 0.11 & 0.06 & 0.24 & 0.29 \\ 
			& & 30 & 0.14 & 0.23 & 0.15 & 0.32 & 0.36 \\ 
			& & 50 & 0.25 & 0.45 & 0.25 & 0.48 & 0.51 \\ 
			& & 100 & 0.46 & 0.73 & 0.38 & 0.71 & 0.74 \\ 
			\hline
		\end{tabular*}
		\begin{tablenotes}
			\item Each entry was based on 1000 replicates. $\rm{WMD_{s}}$  and $\rm{WMD_o}$ were calculated based on 1500 bootstrap samples. Significant level $\alpha$ was set to 0.05.
		\end{tablenotes}
	\end{table}
\end{center}

\subsection{Discrete interval data}
We assessed the performance of $\rm {WMD_o}$ for analyzing discrete interval data, and compared to several methods, including $\rm{W_{cp}}$, $\rm{W_{im}}$, $\rm {WMD_s}$ and cumulative link mixed model (CLMM). Note that we did not include t-test and LMM in the comparison because those methods were designed for continuous data. Instead, we included CLMM in the comparison, which is more appropriate for discrete data. For CLMM, we used {\it clmm} function in the {\it ordinal} R-package\cite{christensen2019tutorial}. The {\it nAGQ} option in {\it clmm} function was set to 5 for best performance as suggested by the authors. Our data were simulated by random sampling with replacement from the complete pairs data of PAR4 from the breast cancer recurrence study described in section 4.2. We considered two correlation scenarios of the paired data. The first scenario adopted the correlation from the real data, which had a  Spearman correlation coefficient $R= 0.49$. The second scenario considered a higher correlation, where we increased the paired data correlation using the following steps. First, we calculated the 40 and 60 percentiles of Group 1, denoted by $Q_1$ and $Q_2$, respectively. Second, for each observation $O$ in Group 2, if $O< Q_1$, we decreased its value by one level; if $O\ge Q_2$, we increased its value by one level. The resulting data had a correlation coefficient $R=0.74$. Other aspects of the simulation setting were the same as those for continuous data described in the previous subsection.

Table ~\ref{par4_type1power} shows the type I error rate for each method. $\rm {WMD_o}$ had a type I error rate close to the nominal value in all cases. $\rm{W}_{\rm {cp}}$ was slightly conservative when the sample size was small. $\rm {WMD_s}$ and CLMM had slightly inflated type I error rate for a few small sample size cases.  $\rm{W}_{\rm {im}}$ had substantially inflated type I error rate in all scenarios. Table ~\ref{par4_type1power} also shows the results for power comparison. We again excluded $\rm{W}_{\rm{im}}$ due to its substantially inflated Type I error rates. $\rm {WMD_s}$ had the best performance across all scenarios. The power of $\rm {WMD_o}$ was higher than that of CLMM, especially for small sample sizes. $\rm {WMD_o}$ also had higher power than $\rm {WMD_s}$, and the difference became larger as the correlation $R$ increased from 0.49 to 0.74.
\begin{center}
	 \begin{table}[h!]%
		\caption{Simulation results for type I error rates and power comparison based on discrete interval data\label{par4_type1power}}%
			\hspace{1cm}{\small\begin{tabular*}{500pt}{lccccccccccccccccc}
			\cline{1-13}
			\multirow{2}{*}{\textbf{Corr}} & \multirow{2}{*}{\textbf{Missing}} &
			\multirow{2}{*}{\textbf{n}} &
			\multicolumn{5}{c}{\textbf{Type I error}} & &
			\multicolumn{4}{c}{\textbf{Power}} \\
			\cline{4-8} \cline{10-13}
			& &	 & $\mathbf{W_{cp}}$  & $\mathbf{W_{im}}$ & \textbf{CLMM} & $\mathbf{WMD_s}$  & $\mathbf{WMD_o}$ & & $\mathbf{W_{cp}}$   & \textbf{CLMM} & $\mathbf{WMD_s}$  & $\mathbf{WMD_o}$ \\
			\cline{1-13}
			0.75 & (0\%, 50\%) & 20 & 0.02 & 0.14 & 0.04 & 0.06 & 0.05 & &  0.16 & 0.19 & 0.24 & 0.28 \\ 
			& & 30 & 0.03 & 0.14 & 0.05 & 0.05 & 0.05 & & 0.32 & 0.30 & 0.34 & 0.39 \\
			& & 50 & 0.04 & 0.16 & 0.06 & 0.05 & 0.05 & & 0.58 & 0.47 & 0.49 & 0.62 \\ 
			& & 100 & 0.06 & 0.21 & 0.08 & 0.05 & 0.06 & & 0.92 & 0.83 & 0.77 & 0.94 \\ 
			& (50\%, 50\%) & 20 & 0.00 & 0.35 & 0.06 & 0.11 & 0.07 & & 0.02 & 0.08 & 0.21 & 0.21 \\ 
			& & 30 & 0.01 & 0.40 & 0.04 & 0.07 & 0.05 & & 0.08 & 0.14 & 0.23 & 0.26 \\ 
			& & 50 & 0.01 & 0.44 & 0.04 & 0.06 & 0.03 & & 0.24 & 0.26 & 0.24 & 0.40 \\ 
			& & 100 & 0.04 & 0.39 & 0.06 & 0.06 & 0.05 & & 0.60 & 0.57 & 0.41 & 0.74 \\ 
			0.50 & (0\%, 50\%) & 20 & 0.03 & 0.12 & 0.04 & 0.06 & 0.06 & & 0.11 & 0.19 & 0.24 & 0.25 \\ 
			& & 30 & 0.04 & 0.14 & 0.05 & 0.07 & 0.07 & & 0.26 & 0.35 & 0.35 & 0.38 \\ 
			& & 50 & 0.04 & 0.16 & 0.04 & 0.05 & 0.05 & & 0.41 & 0.53 & 0.51 & 0.54 \\ 
			& & 100 & 0.06 & 0.25 & 0.04 & 0.05 & 0.05 & &  0.71 & 0.83 & 0.78 & 0.83 \\ 
			& (50\%, 50\%) & 20 & 0.01 & 0.31 & 0.05 & 0.09 & 0.07 & & 0.02 & 0.12 & 0.21 & 0.22 \\ 
			& & 30 & 0.01 & 0.39 & 0.05 & 0.09 & 0.06 & & 0.10 & 0.20 & 0.25 & 0.27 \\
			& & 50 & 0.03 & 0.43 & 0.03 & 0.05 & 0.04 & & 0.17 & 0.32 & 0.35 & 0.36 \\ 
			& & 100 & 0.06 & 0.48 & 0.04 & 0.04 & 0.06 & &  0.39 & 0.66 & 0.54 & 0.68 \\ 
			\cline{1-13}
		\end{tabular*}}
		\begin{tablenotes}
			\item Each entry was based on 1000 replicates. $\rm{WMD_{s}}$  and $\rm{WMD_o}$ were calculated based on 1500 bootstrap samples. Significant level $\alpha$ was set to 0.05.
		\end{tablenotes}
	\end{table}
\end{center}

\section{Examples}
\subsection{A study on prostate cancer biomarkers}
We considered a prostate cancer biomarker study. Data were obtained on 18 patients who underwent surgery for prostate cancer diagnosis from 2011 to 2015 at the University of Kentucky Medical Center and were also consented to provide longitudinal samples of the following biomarkers: PSA, CRP, and IL8.  These samples were obtained at baseline (prior to surgery), 3-month and 6-month follow-ups, and the biomarker levels were measured as continuous variables.  One important question of interest was to compare the pre- and post-surgery levels of those biomarkers. All patients had those biomarkers recorded at baseline, 61\% had missing values at 3-month and 50\% had missing values at 6-month due to missed visits or loss of follow-up.  

We applied our proposed method as well as existing methods to compare PSA, CRP, and IL8 levels at 3-month and 6-month follow-ups to the baseline.  One patient was identified as an outlier based on the Dixon test \cite{komsta2011package} for each of the three biomarkers, and was excluded from the  analysis. Table \ref{tab:PSA} shows p-values based on $\rm{t}_{\rm {cp}}$, LMM,  $\rm{W}_{\rm {cp}}$, $\rm{WMD}_{\rm {s}}$, and $\rm{WMD}_{\rm {o}}$. For the 3-month follow-up vs. baseline comparison, none of the methods had any significant result for any of the three biomarkers. For the 6-month follow-up vs. baseline comparison of PSA measurement, all methods were significant at 0.05 level. Further, $\rm{WMD}_{\rm {s}}$ and $\rm{WMD}_{\rm {o}}$ were also significant at 0.001 level. For the 6-month follow-up vs. baseline comparison of CRP measurement, none of the method yielded a significant result. For the 6-month follow-up vs. baseline comparison of IL8 measurement, $\rm{t}_{\rm {cp}}$, LMM,  $\rm{W}_{\rm {cp}}$, and $\rm{WMD}_{\rm {o}}$ were all significant at 0.05 level while $\rm{WMD}_{\rm {o}}$ was not significant. Further, $\rm{WMD}_{\rm {o}}$ was also significant at 0.001 level.

\begin{center}
	\begin{table}[h!]
		\centering
		\caption{Data analysis results for the prostate cancer study}%
		\begin{tabular*}{500pt}{@{\extracolsep\fill}lcccccccc@{\extracolsep\fill}}
			\hline
			\textbf{Biomarker} & \textbf{Comparison} & $\mathbf{t_{cp}}$    & $\mathbf{LMM}$ & $\mathbf{W_{cp}}$ &  $\mathbf{WMD_s}$ & $\mathbf{WMD_o}$  \\
			\hline
			PSA & 3-month vs baseline &  0.549  & 0.351 & 0.402  & 0.538 & 0.487 \\ 
			& 6-month vs baseline & 0.023 &  0.007 & 0.014 & $<$.001  & $<$.001  \\ 
			CRP & 3-month vs baseline & 0.413  & 0.261 & 0.834  & 0.051 & 0.120 \\ 
			& 6-month vs baseline & 0.344  & 0.384 & 0.554 & 0.295 & 0.230 \\ 
			IL8 & 3-month vs baseline & 0.615  & 0.739 & 0.800  & 0.731 & 0.778 \\ 
			& 6-month vs baseline & 0.013 & 0.015 & 0.024  & 0.756 & 0.006 \\ 
			\hline
		\end{tabular*}\label{tab:PSA}
		\begin{tablenotes}
			\item Numbers presented are p-values. P-values for $\rm{WMD_s}$  and $\rm{WMD_o}$ were calculated based on 1500 bootstrap samples. 
		\end{tablenotes}
	\end{table}
\end{center}

\subsection{A study on biomarkers for breast cancer recurrence}
The study \cite{he2019potential} evaluated expressions of five key cancer-related proteins, including PAR4, TWIST, pCrk/CrkL, SNAIL, and $\beta$-catenin, in paired primary and recurrent tumor tissue specimens from 22 patients that were diagnosed and treated at University of Kentucky Medical Center between 2000 and 2007. Two representative
tissue cores were taken from each tumor and included in a tissue microarray construction. Based on immunohistochemistry, each tissue core was
assigned an intensity score of 0 (negative), 1 (weak), 2 (moderate), or 3 (strong), and the  scores of the two cores for each tumor were averaged. Due to the lack of remaining tissue, lack of invasive carcinoma cells or difficulty in interpreting tissues, measurements of some proteins in some tissue samples were missing. The missing proportions are listed in Table  \ref{tab:protein}.

We compared the expression of those proteins between primary and recurrent tumor tissue samples.  Table \ref{tab:protein} shows the results based on ${\rm W}_{\rm{cp}}$, CLMM,  $\rm{WMD}_{\rm {s}}$, and $\rm{WMD}_{\rm {o}}$. For PAR4, $\rm{WMD}_{\rm {s}}$ and $\rm{WMD}_{\rm {o}}$ identified a significant difference between primary and recurrent tumors, while the other two methods gave non-significant results. For TWIST, CLMM, $\rm{WMD}_{\rm {s}}$ and $\rm{WMD}_{\rm {o}}$ were significant, while ${\rm W_{cp}}$ was not significant. For pCrk/CrkL, all methods were significant. For SNAIL and $\beta$-catenin, none of the methods gave significant results.
\begin{center}
	\begin{table}[h!]
		\centering
		\caption{Data analysis results for the breast cancer study}%
		\begin{tabular*}{500pt}{@{\extracolsep\fill}lccccc@{\extracolsep\fill}}
			\hline
			\textbf{Protein biomarker} &  \textbf{Missing} & $\mathbf{W^*_{\rm {cp}}}$  &\textbf{CLMM} & $\mathbf{WMD_s}$  & $\mathbf{WMD_o}$ \\
			\hline
			PAR4 & (5\%, 0\%) & 0.112 & 0.061 & 0.041 & 0.043 \\ 
			TWIST & (5\%, 19\%) & 0.063 & 0.038 & 0.008 & 0.016  \\ 
			pCrk/CrkL & (14\%, 0\%) & 0.001 &  $<$.001 & $<$.001 & $<$.001 \\ 
			SNAIL & (14\%, 5\%) & 0.634 & 0.419 & 0.355 & 0.484 \\ 
			$\beta$-catenin &  (5\%, 5\%) & 0.255  & 0.317 & 0.256 & 0.235 \\ 
			\hline
		\end{tabular*}\label{tab:protein}
		\begin{tablenotes}
			\item The missing proportions in parenthesis are for primary and recurrent tumor tissue samples, respectively. Other numbers presented are p-values. *: P-values based on the $\rm{W_{\rm {cp}}}$ test have been previously reported in He \textsl{et al.} (2019) \cite{he2019potential}. P-values for $\rm{WMD_s}$  and $\rm{WMD_o}$ were calculated based on 1500 bootstrap samples. 
		\end{tablenotes}
	\end{table}
\end{center}

\section{Discussion}
We have developed a class of test statistics based on the difference in weighted sample means between two groups for partially paired data. This class of tests is very general because it allows different weights for observations from different groups as well as from complete and incomplete pairs within the same group. We have further derived the optimal weight for this class of tests. Simulation results demonstrate that our test with the optimal weight is more powerful than existing methods.

Our tests are free of any model or distributional assumptions. The calculation of tests is direct and does not require any modeling of the correlation between the paired outcomes. In addition, the p-value is calculated based on the asymptotic normality of the test statistics which does not require any assumption on the distribution of the data. As they do not impose any distributional assumption, our tests provide a unified approach to deal with both continuous and discrete interval data. 

There are two strategies to develop tests specifically for partially paired data. The first strategy considers weighted linear combinations of test statistics comparing paired data and test statistics comparing unpaired data \cite{dubnicka2002rank, brunner2002nonparametric, konietschke2012ranking, fong2017rank, bhoj1978testing}. The second strategy first constructs estimators of the population mean difference between the two groups, and then derives test statistics based on the estimators\cite{martinez2013hypothesis}. An advantage of the second strategy is that in addition to the p-value, it provides estimates for the population mean difference between groups, quantifying the magnitude of the difference in a meaningful way. We take the second strategy in this paper and develop a general class of estimators for the population mean difference. We further obtain the efficient estimator by deriving the optimal weight that minimizes variances of estimators within the class.

Like other methods, our tests require the MCAR assumption. Here we consider the situation that there is no additional information collected except for the paired outcomes. When additional information on covariates $X$ is available and the data is missing at random (MAR) given $X$, our tests can be extended to such situation by weighting each observation by the inverse of the estimated propensity score, i.e. the probability of the corresponding outcome is observed given the $X$ value of the corresponding individual. However, the derivation of the asymptotic variance of the extended tests and the corresponding optimal weight will be more complex.


\section*{Acknowledgments}
This research was partially supported by the National Institute of Health (1R03CA179661-01A1, 5P20GM103436-15) and the Biostatistics and Bioinformatics Shared Resource Facility of the
University of Kentucky Markey Cancer Center (P30CA177558). The prostate cancer biomarker study was supported by the National Cancer Institute (5R01CA143428). We thank Drs. Tom Tucker, Rina Plattner, Vivek Rangnekar, Binhua Zhou and Chunming Liu for sharing their data on protein biomarkers for breast cancer recurrence.

\appendix

\section{Proof of Theorem 1.}
\label{wmd_var}
In this section, we show the asymptotic normality of the estimator $\hat{D}(\hat{w}_1, \hat{w}_2)$ using modern empirical process theory. Let $\CPn$ and $\CP$ denote the empirical measure
and the distribution under the true model, respectively. For a measurable function $f$ and measure $Q$, the integral $\int f dQ$ is abbreviated as $Qf$. We consider the following three situations.
\\

\noindent\textbf{Situation 1: Missing values present in both groups.}\\
First,
\begin{eqnarray}\label{equation1}
	\sqrt{n}\left\{\hat{D}(\hat{w}_1, \hat{w}_2) - (\mu_1 - \mu_2)\right\} &=\sqrt{n}\left\{\hat{D}({w}_1, {w}_2) - (\mu_1 - \mu_2)\right\}+\sqrt{n}\left\{\hat{D}(\hat{w}_1, \hat{w}_2) - \hat{D}({w}_1, t{w}_2) \right\}.
\end{eqnarray}
Under MCAR assumption, the first term of (\ref{equation1}) can be written as 
\begin{eqnarray*}
	\sqrt{n}\left\{\frac{\CPn f_1}{{\CPn} g_1}-\frac{\CP f_1}{{\CP} g_1}\right\}
	-\sqrt{n}\left\{\frac{\CPn f_3}{{\CPn} g_3}-\frac{\CP f_3}{{\CP} g_3}\right\}
	+\sqrt{n}\left\{\frac{\CPn f_2}{{\CPn} g_2}-\frac{\CP f_2}{{\CP} g_2}\right\}
	-\sqrt{n}\left\{\frac{\CPn f_4}{{\CPn} g_4}-\frac{\CP f_4}{{\CP} g_4}\right\}.
\end{eqnarray*}
where 
\begin{eqnarray*}
	&&f_1 =w_1Y_1R_1R_2,\,\, g_1=R_1R_2,\\
	&&f_2 =(1-w_1)Y_1R_1(1-R_2),\,\, g_2=R_1(1-R_2),\\
	&&f_3 =w_2Y_2R_1R_2,\,\, g_3=g_1=R_1R_2,\\
	&&f_4 =(1-w_2)Y_2R_2(1-R_1),\,\, g_4=R_2(1-R_1).
\end{eqnarray*}
Since $\CPn f_i/\CPn g_i\rightarrow \CP f_i/\CP g_i$, $Pf_i<\infty$ and $g_i$ are bounded monoton functions, there exists a $\CP$-Donsker class $\mathcal{F}$ such that $P^n\left(\frac{g_i\CPn f_i}{P g_i\CPn g_i}\in \mathcal{F}\right)\rightarrow 1$ based on Example 2.10.27 of Van Der Vaart and Wellner (1996) \cite{van1996weak}. In addition, it is easy to verify that  $f_i$, $g_i$ satisfy the other two conditions in Lemma ~\ref{prop_var}. Therefore, based on Lemma ~\ref{prop_var}, we obtain that the first term of (\ref{equation1}) is equal to $\sqrt{n}(\CPn-\CP)\xi+o_p(1)$, where
\begin{eqnarray*}
\xi=\left\{\frac{f_1}{{P}_0g_1}-\frac{g_1{P}_0f_1}{({P}_0g_1)^2} +\frac{f_2}{{P}_0g_2}-\frac{g_2{P}_0f_2}{({P}_0g_2)^2}\right\}-\left\{\frac{f_3}{{P}_0g_3}-
	\frac{g_3{P}_0f_3}{({P}_0g_3)^2} +\frac{f_4}{{P}_0g_4}-\frac{g_4{P}_0f_4}{({P}_0g_4)^2} \right\}.
\end{eqnarray*}
The second term of (\ref{equation1}) can be written as $\sqrt{n}(\hat{w}_1 -w_1)(\bar{Y}_{1}^C -\bar{Y}_{1}^I)-\sqrt{n}(\hat{w}_2 -w_2)(\bar{Y}_{2}^C -\bar{Y}_{2}^I)$, and thus is $o_p(1)$ by the MCAR assumption and the condition that $\sqrt{n}(\hat{w}_1-w_1)$ and $\sqrt{n}(\hat{w}_2-w_2)$ are $O_p(1)$.
Therefore, $\sqrt{n}\left\{\hat{D}(\hat{w}_1, \hat{w}_2) - (\mu_1 - \mu_2)\right\}$ is asymptotically normal with mean 0 and variance   
\begin{eqnarray*}
	\sigma_D^2(w_1, w_2) &=& \mathrm{Var}(\xi)\\
	&=&  \mathrm{Var}\left[ \left\{\frac{w_1R_1R_2}{p_{12}} + \frac{(1-w_1)R_1(1-R_2)}{p_1-p_{12}}\right\}(Y_1 - \mu_1)-
	\left\{\frac{w_2R_1R_2}{p_{12}} +\frac{(1-w_2)R_2(1-R_1)}{p_2-p_{12}}\right\}(Y_2 - \mu_2)\right]\\
	&=& \mathrm{E}\left\{\frac{w_1R_1R_2}{p_{12}} + \frac{(1-w_1)R_1(1-R_2)}{p_1-p_{12}}\right\}^2 \mathrm{Var}(Y_1) + \mathrm{E}\left\{\frac{w_2R_1R_2}{p_{12}} +\frac{(1-w_2)R_2(1-R_1)}{p_2-p_{12}}\right\}^2 \mathrm{Var}(Y_2 )\\
	&& - 2\mathrm{E}\left[ \left\{\frac{w_1R_1R_2}{p_{12}} + \frac{(1-w_1)R_1(1-R_2)}{p_1-p_{12}}\right\}\left\{\frac{w_2R_1R_2}{p_{12}} +\frac{(1-w_2)R_2(1-R_1)}{p_2-p_{12}}\right\}\right]\mathrm{Cov}(Y_1, Y_2) \\
	&=& \left\{\frac{w_1^2}{p_{12}} + \frac{(1-w_1)^2}{p_1 - p_{12}}\right\}\sigma_1^2 +  \left\{\frac{w_2^2}{p_{12}} + \frac{(1-w_2)^2}{p_2 - p_{12}}\right\}\sigma_2^2 - \frac{2w_1w_2\rho \sigma_1 \sigma_2}{p_{12}}.
\end{eqnarray*}	
\\
\noindent\textbf{Situation 2: Missing values present only in Group 1.}\\
When missing values present only in Group 1, we have $w_1=1$, $p_2=1$, $p_{12} = p_1$ and $f_2=0$. Then the first term of (\ref{equation1}) can be written as 
\begin{eqnarray*}
	\sqrt{n}\left\{\frac{\CPn f_1}{{\CPn} g_1}-\frac{\CP f_1}{{\CP} g_1}\right\}
	-\sqrt{n}\left\{\frac{\CPn f_3}{{\CPn} g_3}-\frac{\CP f_3}{{\CP} g_3}\right\}
	-\sqrt{n}\left\{\frac{\CPn f_4}{{\CPn} g_4}-\frac{\CP f_4}{{\CP} g_4}\right\}.
\end{eqnarray*}
Similar to Situation 1, we obtain $\sqrt{n}\left\{\hat{D}(\hat{w}_1, \hat{w}_2) - (\mu_1 - \mu_2)\right\}$ is asymptotically normal with mean 0 and variance   
\begin{eqnarray*}
		\sigma_D^2(w_1, w_2) &= &\mathrm{Var} \left[\left\{\frac{f_1}{{P}_0g_1}-\frac{g_1{P}_0f_1}{({P}_0g_1)^2} \right\}-\left\{\frac{f_3}{{P}_0g_3}-
	\frac{g_3{P}_0f_3}{({P}_0g_3)^2} +\frac{f_4}{{P}_0g_4}-\frac{g_4{P}_0f_4}{({P}_0g_4)^2} \right\}\right]\\
	&=&  \mathrm{Var}\left[ \left\{\frac{R_1R_2}{p_{1}} \right\}(Y_1 - \mu_1)-
	\left\{\frac{w_2R_1R_2}{p_{1}} +\frac{(1-w_2)R_2(1-R_1)}{p_2-p_{1}}\right\}(Y_2 - \mu_2)\right]\\
	&=& \mathrm{E}\left\{\frac{R_1R_2}{p_{1}} \right\}^2 \mathrm{Var}(Y_1) + \mathrm{E}\left\{\frac{w_2R_1R_2}{p_{1}} +\frac{(1-w_2)R_2(1-R_1)}{p_2-p_{1}}\right\}^2 \mathrm{Var}(Y_2 )\\
	&& - 2\mathrm{E}\left[ \left\{\frac{R_1R_2}{p_{1}} \right\}\left\{\frac{w_2R_1R_2}{p_{1}} +\frac{(1-w_2)R_2(1-R_1)}{p_2-p_{1}}\right\}\right]\mathrm{Cov}(Y_1, Y_2) \\
	&=& \frac{\sigma_1^2}{p_{1}} +  \left\{\frac{w_2^2}{p_{1}} + \frac{(1-w_2)^2}{1 - p_{1}}\right\}\sigma_2^2 - \frac{2w_2\rho \sigma_1 \sigma_2}{p_{1}}.\\
\end{eqnarray*}
\\
\noindent\textbf{Situation 3: Missing values present only in Group 2.}\\
When missing values present only in Group 2, we have $w_2=1$, $p_1=1$, $p_{12}=p_2$ and $f_4=0$. Similarly, we obtain $\sqrt{n}\left\{\hat{D}(\hat{w}_1, \hat{w}_2) - (\mu_1 - \mu_2)\right\}$ is asymptotically normal with mean 0 and variance 
$$\sigma_D^2(w_1, w_2) = \left\{\frac{w_1^2}{p_{2}} + \frac{(1-w_1)^2}{1 - p_{2}}\right\}\sigma_1^2 + \frac{\sigma_2^2}{p_{2}} - \frac{2w_1\rho \sigma_1 \sigma_2}{p_{2}}. $$

\hfill $\square$
\\
\begin{lemma}
	\label{prop_var}
	For any measurable functions $f$ and $g$, if there exists a $\CP$-Donsker class $\mathcal{F}$ such that $P^n\left(\frac{g\CPn f}{P g\CPn g}\in \mathcal{F}\right)\rightarrow 1$, $P\left\{\frac{g^2}{(P g)^2}\left (\frac{\CPn f}{\CPn g}-\frac{P f}{P g}\right)^2\right\}\rightarrow 0$, and $\frac{g P f}{(P g)^2}\in L_2(P)$, then
	$$\sqrt{n}\left\{\frac{\CPn f}{\CPn g}- \frac{\CP f}{\CP g}\right\}= \sqrt{n}(P_n-P_0)\left\{ \frac{f}{\CP g} - 
	\frac{g\CP f}{(\CP g)^2} \right\} + o_p(1). $$
\noindent \textsl{Proof.}
	\begin{eqnarray}\label{lemma1}
		&& \sqrt{n}\left\{\frac{\CPn f}{\CPn g}- \frac{\CP f}{\CP g}\right\}\notag\\
		&=& \frac{ \sqrt{n}(\CPn-\CP)f}{\CP g} - \sqrt{n} (\CPn-\CP)g \frac{\CPn f}{\CPn g\CP g}\notag\\
		&=& \sqrt{n}(\CPn-\CP)\left\{\frac{f}{\CP g} - \frac{g\CP f}{(\CP g)^2}  \right\} -
		\sqrt{n}(\CPn-\CP)\left\{  \frac{g\CPn f}{\CPn g\CP g}-\frac{g\CP f}{(\CP g)^2}  \right\}.
	\end{eqnarray}
By the conditions and Lemma 19.24 of Van der Vaart (1998)\cite{van1998asymptotic}, the second term of \ref{lemma1} is $o_p(1)$. 
\hfill $\square$
\end{lemma}
\section{Derivation of the optimal weight}

\noindent\textbf{Situation 1: Missing values present in both groups.}\\
The optimal weight is the $(w_1, w_2)$ values that minimizes the asymptotic variance $\sigma_D^2(w_1, w_2)$. To find the minimization point of  $\sigma_D^2(w_1, w_2)$, we solve the first derivative equations  $\partial \sigma_D^2(w_1, w_2) / \partial w_1 = 0$ and $\partial \sigma_D^2(w_1, w_2) / \partial w_2 = 0$ and obtain the root as
\begin{eqnarray*}
w_{1}^{*} &=&  \frac{p_2p_{12}\sigma_1^2 + \rho\sigma_1\sigma_2p_{12}(p_1-p_{12})}{p_1p_2\sigma_1^2 - \rho^2\sigma_1^2(p_1-p_{12})(p_2-p_{12})},\\
w_{2}^{*} &=&  \frac{p_1p_{12}\sigma_2^2 + \rho\sigma_1\sigma_2p_{12}(p_2-p_{12})}{p_1p_2\sigma_2^2 - \rho^2\sigma_2^2(p_1-p_{12})(p_2-p_{12})}.\\
\end{eqnarray*}
Note that our weight is defined under the constraints that $0 \leq w_1 \leq 1$ and $0 \leq w_2 \leq 1$. If $(w_1^*, w_2^*)$ falls in this squared region, it is the optimal weight. Otherwise, the minimization of the variance will be reached on the boundary of this squared region. Since the boundary consists of four lines, we consider the following set $\mathcal{A}$ as a collection of the minimization point on each of the four lines, 
\begin{eqnarray*}
\mathcal{A}  = & \left \{ \left(0, \frac{p_{12}}{p_2}\right), \left(1, g\left(\frac{p_{12}\sigma_2+\rho\sigma_1(p_2-p_{12})}{p_{2}\sigma_2}\right)\right), \right.\\
& ~~\left.\left(\frac{p_{12}}{p_1}, 0\right), \left(g\left(\frac{p_{12}\sigma_1+\rho\sigma_2(p_1-p_{12})}{p_{1}\sigma_1}\right),1\right)\right\},\\
\end{eqnarray*}
where
\begin{displaymath}
g(z) = \left\{
\begin{split}
0 & \hspace{0.3cm}  \textrm{if } z < 0,\\
z & \hspace{0.3cm}  \textrm{if } 0 \leq z \leq 1,\\
1 & \hspace{0.3cm}  \textrm{if } z > 1.\\
\end{split}\right.
\end{displaymath}
Then optimal weight is chosen as the point in $\mathcal{A}$ that minimizes the variance. In summary, the optimal weight is
\begin{eqnarray*}\label{eq:weight}
(w_1^{opt}, w_2^{opt}) = \left\{
\begin{split}
(w_1^*, w_2^*) \hspace{1cm} & \hspace{0.3cm}  \textrm{if } 0 \leq w_1^* \leq 1, 0 \leq w_2^* \leq 1,\\
\underset{(w_1, w_2) \in \mathcal{A}}{\textrm{argmin}} \sigma_D^2(w_1, w_2) & \hspace{0.3cm}  \textrm{otherwise}. \\
\end{split}.\right.
\end{eqnarray*}
\\


\noindent\textbf{Situation 2: Missing values present only in Group 1.}\\
When missing values only present in group 1, we have $p_2=1$, $p_{12}=p_1$ and $w_1=1$. The optimal weight of $w_2$ is the point that minimizes $\sigma_D^2(1, w_2) $. We solve $d \sigma_D^2(1, w_2) / d w_2 = 0$ and obtain the root as $w_{2}^{*} = \frac{p_{1}\sigma_2 +\rho\sigma_1 (1-p_{1})}{\sigma_2}$. Note  that $0 \leq w_2 \leq 1$. If $0 \leq w_2^* \leq 1$, the optimal weight is $w_2^*$. If $ w_2^* < 0$, the optimal weight is 0.  If $w_2^* > 1$, the optimal weight is 1. In summary, the optimum weight of $w_2$ is $w_2^{opt} = g(w_2^*)$. Similarly, when missing values only present in group 2, we have $p_1=1$, $p_{12}=p_2$ and $w_2=1$. The optimum weight of $w_1$ is $w_1^{opt} = g(w_1^*)$, where $w_{1}^{*} = \frac{p_{2}\sigma_1 +\rho\sigma_2 (1-p_{2})}{\sigma_1}$.

As shown in Situation 1, $\sigma_D^2(w_1, w_2)$ is minimized at  $w_{1}^{*} = \frac{p_{2}\sigma_1 +\rho\sigma_2 (1-p_{2})}{\sigma_1}$. Under the constraint that $0 \leq w_2 \leq 1$, the optimum weight is $w_1^{opt} = g(w_1^*)$ .

When missing values present only in Group 1, we have
\begin{align*}
	\sigma_D^2(w_1, w_2) &=  \frac{1}{p_{12}}\sigma_1^2 + \frac{w_2^2 - 2p_{1}w_2 + p_{1}}{p_{1}(1-p_{1})}\sigma_2^2 - \frac{2w_2\rho\sigma_1\sigma_2}{p_{1}}\\
	& = w_2^2\frac{\sigma_2^2}{p_{1}(1-p_{1})} -2 w_2\left\{ \frac{p_{1}\sigma_2^2}{p_{1}(1-p_{1})} + \frac{\rho\sigma_1\sigma_2}{p_{1}} \right\} + C_4
\end{align*}
where $C_4$ is a constant. Then $\sigma_D^2(w_1, w_2)$ is minimized at 
\begin{align*}
	w_2^* & = \left\{ \frac{p_{1}\sigma_2^2}{p_{1}(1-p_{1})} + \frac{\rho\sigma_1\sigma_2}{p_{1}} \right\}/ \left\{\frac{\sigma_2^2}{p_{1}(1-p_{1})} \right\}\\
	& = \frac{p_{1}\sigma_2 +\rho\sigma_1 (1-p_{1})}{\sigma_2}
\end{align*}
\\
\noindent\textbf{Situation 3: Missing values present only in Group 2.}\\
Similarly, when missing values present only in Group 2, webtain the optimal weight as 
\[w_1^*  =  \frac{p_{2}\sigma_1 +\rho\sigma_2 (1-p_{2})}{\sigma_1}.\]

\hfill $\square$

\nocite{*}
\bibliography{WMD-refs}

\begin{thebibliography}{10}
\providecommand \doibase [0]{http://dx.doi.org/}%

\bibitem{he2019potential}
He C, Plattner R, Rangnekar V, et al. Potential protein markers for breast
  cancer recurrence: a retrospective cohort study. {\it Cancer Causes \&
  Control} 2019\string; 30(1)\string: 41--51.

\bibitem{little2002statistical}
Little RJ, Rubin DB. {\it Statistical analysis with missing data}. 793.
\newblock John Wiley \& Sons .
\newblock 2019.

\bibitem{dubnicka2002rank}
Dubnicka SR, Blair RC, Hettmansperger TP. Rank-based procedures for mixed
  paired and two-sample designs. {\it Journal of Modern Applied Statistical
  Methods} 2002\string; 1(1)\string: 6.

\bibitem{brunner2002nonparametric}
Brunner E, Domhof S, Langer F, Brunner E. {\it Nonparametric analysis of
  longitudinal data in factorial experiments}.
\newblock J. Wiley New York .
\newblock 2002.

\bibitem{konietschke2012ranking}
Konietschke F, Harrar SW, Lange K, Brunner E. Ranking procedures for matched
  pairs with missing data$-$asymptotic theory and a small sample approximation.
  {\it Computational Statistics \& Data Analysis} 2012\string; 56(5)\string:
  1090--1102.

\bibitem{fong2017rank}
Fong Y, Huang Y, Lemos MP, Mcelrath MJ. Rank-based two-sample tests for paired
  data with missing values. {\it Biostatistics} 2017\string; 19(3)\string:
  281--294.

\bibitem{bhoj1978testing}
Bhoj DS. Testing equality of means of correlated variates with missing
  observations on both responses. {\it Biometrika} 1978\string; 65(1)\string:
  225--228.

\bibitem{martinez2013hypothesis}
Mart{\'\i}nez-Camblor P, Corral N, Hera M.~d.~lJ. Hypothesis test for paired
  samples in the presence of missing data. {\it Journal of Applied Statistics}
  2013\string; 40(1)\string: 76--87.

\bibitem{nlme2019}
Pinheiro J, Bates D, DebRoy S, Sarkar D, {R Core Team} . {\it {nlme}: Linear
  and Nonlinear Mixed Effects Models}. R Foundation for Statistical Computing;
  Vienna, Austria:  2021.
\newblock R package version 3.1-153.

\bibitem{christensen2019tutorial}
Christensen RHB. A Tutorial on fitting Cumulative Link Mixed Models with clmm2
  from the ordinal Package.
  \url{https://cran.r-project.org/web/packages/ordinal/vignettes/clmm2_tutorial.pdf};
  2019.

\bibitem{komsta2011package}
Komsta L. {\it {outliers}: Tests for outliers.\,\,}. R Foundation for
  Statistical Computing; Vienna, Austria:  2015.
\newblock R-package version 0.14.

\bibitem{van1996weak}
Van Der~Vaart AW, Wellner J. {\it Weak convergence and empirical processes:
  with applications to statistics}.
\newblock Springer Science \& Business Media .
\newblock 1996.

\bibitem{van1998asymptotic}
Vaart V.~dAW. {\it Asymptotic statistics}. 3.
\newblock Cambridge university press .
\newblock 1998.

\bibitem{lin1971estimation}
Lin PE. Estimation procedures for difference of means with missing data. {\it
  Journal of the American Statistical Association} 1971\string; 66(335)\string:
  634--636.

\bibitem{bates2014fitting}
Bates D, M{\"a}chler M, Bolker B, Walker S. Fitting linear mixed-effects models
  using lme4. {\it arXiv preprint arXiv:1406.5823} 2014.

\end{thebibliography}
\bibliographystyle{WileyNJD-AMA}

\end{document}